\documentclass[12pt]{article}
\usepackage{graphicx}
\usepackage{natbib}

\newcommand\pubnumber{}
\newcommand\pubdate{\today}

\def\Title#1{\begin{center} {\Large #1 } \end{center}}
\def\Author#1{\begin{center}{ \sc #1} \end{center}}
\def\Address#1{\begin{center}{ \it #1} \end{center}}

\newcommand\pubblock{\rightline{\begin{tabular}{l} \pubnumber\\
         \pubdate  \end{tabular}}}
\newenvironment{Abstract}{\begin{quotation}  }{\end{quotation}}
\newenvironment{Presented}{\begin{quotation} \begin{center} 
             PRESENTED AT\end{center}\bigskip 
      \begin{center}\begin{large}}{\end{large}\end{center} \end{quotation}}
\def\Acknowledgements{\bigskip  \bigskip \begin{center} \begin{large}
             \bf ACKNOWLEDGEMENTS \end{large}\end{center}}

\def\beq{\begin{equation}}
\def\eeq#1{\label{#1}\end{equation}}
\def\eeqn{\end{equation}}

\def\beqa{\begin{eqnarray}}
\def\eeqa#1{\label{#1}\end{eqnarray}}
\def\eeqan{\end{eqnarray}}

\let\bar=\overbar

\def\Dslash{\not{\hbox{\kern-4pt $D$}}}
\def\dslash{\not{\hbox{\kern-2pt $\del$}}}

\def\msb{{\bar{\ssstyle M \kern -1pt S}}}

\begin{document}
\begin{titlepage}
\pubblock

\vfill
\Title{Current Status of Very-Large-Basis Hamiltonian Diagonalizations for Nuclear Physics}
\vfill
\Author{Calvin W. Johnson\footnote{This material is based upon work supported by the U.S. Department of Energy, 
Office of Science, Office of Nuclear Physics, under Award Number  DE-FG02-03ER41272}}
\Address{Department of Physics\\
San Diego State University, 92182-1233, CA, USA}
\vfill
\begin{Abstract}
Today there are a plethora of many-body techniques for calculating nuclear wave functions and matrix elements. I review the status of that reliable workhorse, the interacting shell model, a.k.a. configuration-interaction methods, a.k.a. Hamiltonian diagonalization, and survey its advantages and disadvantages.  With modern supercomputers one can tackle dimensions up to about 20 billion! I  discuss how we got 
there and where we might go in the near future.
\end{Abstract}
\vfill
\begin{Presented}
Thirteenth Conference on the Intersections of Particle and Nuclear Physics\\
Indian Wells, CA,  May 28--June 3, 2018
\end{Presented}
\vfill
\end{titlepage}
\def\thefootnote{\fnsymbol{footnote}}
\setcounter{footnote}{0}

\section{Introduction and relevance of nuclear structure}

Some of my colleagues think  nuclear structure theory, in particular the nuclear shell model, is as old-fashioned as the horse-and-buggy.  
But  really it's is the exact opposite. Nuclear structure theory has lots of exciting  developments.  These developments push the shell model from 
phenomenology to rigorous first-principle calculations,  driven partly by new ideas but above all by the explosion in computing 
capabilities. While in the early days solving a $25 \times 25$ matrix was the height of computation \citep{PhysRev.105.1563}, today we find extremal eigenvalues 
of matrices exceeding dimensions of $2 \times 10^{10}$ \citep{PhysRevC.97.034328}.

Aside from the intrinsic physics interest of nuclei,  careful microscopic calculations are needed for many applications.  Detection of known and unknown particles, 
from neutrinos \citep{PhysRevC.74.034307} to dark matter \citep{PhysRevC.89.065501}, as well as experiments testing fundamental symmetries, such as neutrinoless double-$\beta$ decay 
\citep{PhysRevLett.110.222502}
and nonconservation of parity and time-reversal symmetries \citep{haxton2001atomic}, often 
require  knowledge of  matrix elements in complex nuclei.  For such calculations 
to be reliable and both precise and accurate, they need to be founded on solid microscopic calculations.  Fortunately, in many cases 
modern nuclear structure theory is rising to the challenge.

\section{Key ideas  in large-basis diagonalization}

 This paper deals solely with diagonalization of the many-body Hamiltonian in a basis built from shell-model 
single-particle states, also called the 
\textit{configuration-interaction} method or the \textit{interacting shell model} \citep{BG77,br88,ca05}.  The  idea is straightforward:  
expand a state $| \Psi \rangle$ in a basis $\{ | \alpha \rangle \}$ (assumed to be orthonormal, 
$\langle \alpha | \beta \rangle = \delta_{\alpha, \beta }$),
\begin{equation}
| \Psi \rangle = \sum_\alpha  c_\alpha  | \alpha \rangle;
\end{equation}
 minimizing $ \langle \Psi | \hat{H} | \Psi \rangle / \langle \Psi | \Psi \rangle$ leads to the eigenvalue equation
\begin{equation}
\sum_\beta H_{\alpha, \beta} c_\beta = E c_\alpha, \label{mateigen}
\end{equation}
where $H_{\alpha, \beta} = \langle \alpha | \hat{H} | \beta \rangle$ is the matrix element of the many-body Hamiltonian $\hat{H}$ in 
this basis.  I deal with the question of the choice of basis in section \ref{basis}





We can broadly classify configuration-interaction (CI) calculations into two categories, \textit{phemonenological}
 and \textit{ab initio}.  Phenomenological calcqulations are older, and usually 
assume a fixed cored and a relatively narrow valence space, such as the $1s_{1/2}$-$0d_{3/2}$-$0d_{5/2}$ space 
with a fixed $^{16}$O core, or the $1p$-$0f$ space with a fixed $^{40}$Ca core\citep{BG77,br88,ca05}.   The interactions actually start from 
some \textit{ab initio} underlying interaction, and then adjusted to many-body spectra in the target space \citep{PhysRevC.74.034315}. Because of this, 
it is fair to call them \textit{semi}-phenomenological.  By \textit{ab initio}
I mean a potential fitted to few-body data, such as nucleon-nucleon scattering and the binding energies of the $A=2,3$ and 
other light systems. These interactions are most commonly built from chiral effective field theory\citep{PhysRevC.68.041001}, 
but not always \citep{wiringa1995accurate,shirokov2016n3lo}.  
Despite having essentially the same few-body input, difference choices such as cut-off regulators \citep{PhysRevC.94.034001} can strongly influence the 
final many-body energies. 

 Purely \textit{ab initio} CI calculations are often called 
\textit{no-core shell model} (NCSM) calculations \citep{navratil2009recent,barrett2013ab}, precisely because there is no core: all particles, in principle, are active, and 
the standard methodology increases the model space until convergence: see section \ref{converge} below.

In between these two are attempts to derive \textit{ab initio} effective interactions, with no adjustable parameters, for phenomenolo\-gi\-cal-like valences spaces for medium 
and heavy nuclei, via a double projection (Okubo-Lee-Suzuki) 
method \citep{PhysRevC.91.064301}, 
via coupled clusters \citep{PhysRevLett.113.142502}, and via the 
in-medium similarity renormalization group \citep{PhysRevLett.118.032502}.

Because we cast the many-body Schr\"odinger equation as a matrix equation, the main computational task becomes solving a 
matrix eigenvalue problem.
While some bases are larger than other, as discussed below, almost all CI calculations involve large enough dimensions that it would 
be foolish to try to find all eigenpairs.  Instead, one solves for extremal eigenvalues using Arnoldi-type algorithms, almost always 
the Lanczos algorithm \citep{Lanczos}, although there have been attempts to use other methods \citep{shao2018accelerating}.

\section{Basis states for configuration interaction}

\label{basis}

How to  construct the basis set $\{ | \alpha \rangle \}$?  One choice is to use many simple states.  The most common building block 
are  Slater determinants (antisymmeterized products of single-particle states) 
or more generally the occupation-space representations of Slater determinants using creation and annihilation operators.
Furthermore, one often uses an is  \textit{M-scheme} basis, where each Slater determinant has the same fixed total 
$M$ or $J_z$, that is, the $z$-component of angular momentum. This is easy  because $J_z$ is an additive quantum number. 
Many CI shell model codes use an $M$-scheme basis, most notably {\tt ANTOINE} \citep{ANTOINE}, 
{\tt MFDn} \citep{MFDn},  {\tt BIGSTICK} 
\citep{BIGSTICK,johnson2018bigstick}, and {\tt KSHELL} \citep{shimizu2013nuclear}.
$M$-scheme bases are simple, amenable to a bit occupation representation ideal for digital computers \citep{Lanczos}, 
and one can compute matrix elements in the basis efficiently.  The drawback is one needs a large number of $M$-scheme 
basis states to build up nuclear correlations.

There are more sophisticated bases. $J$-scheme basis  states have fixed total angular momentum $J$. 
The most widely used $J$-scheme codes are {\tt OXBASH} \citep{OXBASH} and its successor {\tt NuShellX} \citep{NuShellX}.
As such, the $J$-scheme basis has smaller dimensions than the $M$-scheme. One can go even further, to so-called 
\textit{symmetry-adapted} bases, based upon groups such as SU(3) \citep{PhysRevLett.111.252501} or Sp(3,R) \citep{mccoysymplectic}. When judiciously truncated in the choice of irreps (subspaces defined by the Casimir operators of the group), such calculations can be even smaller in dimension. 

 Dimensions alone do not measure the computational burden.  From Eq.~(\ref{mateigen})  the real computational 
burden is in the nonzero matrix elements of the Hamiltonian.  $M$-scheme bases are very sparse, as small as
$\sim 10^{-6}$, while $J$-scheme bases,  smaller in dimensions, can have \textit{more} nonzero matrix elements, and 
symmetry-adapted bases yet more \citep{dytrych2016efficacy}.  Furthermore, $J$-scheme basis states are generally represented as a linear combination of 
$M$-scheme states, and symmetry-adapted states are either a linear combination of $M$-scheme states or require non-trivial recursion algorithms, making calculation of the nonzero matrix elements a significant burden; by contrast, in the $M$-scheme 
matrix elements are so simple they can be recomputed efficiently on-the-fly, dramatically reducing the memory load, albeit at a price 
of a more complicated algorithm \citep{ANTOINE,johnson2018bigstick}.  There is no `best' 
 basis,
only the recognition of trade-offs. 

In addition to the choice of many-body basis states, there is the question of the underlying single-particle basis.  Phenomenological 
calculations either assume a harmonic oscillator basis or a Woods-Saxon like basis, but in general as matrix elements are primarily 
tuned to spectra, the single-particle basis is ambiguous.  More rigorous \textit{ab initio} calculations such as the no-core shell model 
(NCSM) do have definite singe-particle bases, almost always harmonic oscillator which aids in removing spurious center-of-mass 
motion.  Yet harmonic oscillator wave functions have a steep, unphysical fall off. Hence there have been many efforts to 
introduce better wave functions \citep{PhysRevC.86.034312}, a question which has proved challenging. The most promising seem to be 
\textit{natural orbitals} \citep{constantinou2017natural}, orbitals that diagonalize the ground state one-body 
density matrix.


\section{Convergence and extrapolation}
\label{converge}

Phenomenological calculations take place in a fixed set of valence orbits (unfortunately common usage often conflates 
\textit{orbits} and \textit{shells}), with interactions tuned to that valence space, such as
the $1s$-$0d$ or $sd$-space \citep{PhysRevC.74.034315}.   \textit{Ab initio} calculations, conversely, imply a result in a unrestricted 
or infinite space.  Because any actual calculation must be done in a finite space, one must investigate the convergence as 
the space is increased, and in many cases, extrapolate to the infinite limit. 

In default NCSM calculations \citep{barrett2013ab}, one defines the model space by two parameters: the harmonic oscillator 
frequency $\Omega$, or, more often, $\hbar \Omega$, for the single-particle basis states, and $N_\mathrm{max}$, the maximum 
number of oscillator quanta allowed above the lowest configuration; historically this has also been called $N\hbar \Omega$. 
Typically one wants to extrapolate to infinite $N_\mathrm{max}$ and $\hbar \Omega$. 

One strategy is to use an exponential extrapolation, e.g. fitting energies to a form $a + b \exp (-c N_\mathrm{max})$
 \citep{heng2017ab}. This is inspired by similar exponential extrapolations in phenomenological shell model calculations 
 where even the finite model space is so large one must truncate and extrapolate  \citep{PhysRevLett.82.2064}.  
 For the NCSM, however, the results are not very robust.  Instead,
 recent work has found more robust extrapolation by combining $N_\mathrm{max}$ and $\hbar \Omega$ in to 
infrared and ultraviolet parameters, and following the convergence in 
those parameters \citep{PhysRevC.86.054002,PhysRevC.87.044326,PhysRevC.91.061301}. This can also be linked to interpreting 
$N_\mathrm{max}$ as a finite `wall' \citep{PhysRevC.86.031301}.

In a way, these extrapolations are brute force, and limited by the capability of modern computers. The basis dimension grows 
exponentially with the number of orbits / $N_\mathrm{max}$ and particles, 
which is why size-extensive methods such as coupled clusters \citep{hagen2010ab} are attractive, 
but which have their own set of limitations.   These limitations inspire  alternatives to the standard NCSM prescription: rather 
than brute force computation in a larger basis, build in \textit{smarter} bases, such as use of
better single orbitals such as natural orbitals \citep{constantinou2017natural}, and selected irreducible representations 
in symmetry-adapted bases which efficiently exploit deformation degrees of freedom \citep{PhysRevLett.111.252501,mccoysymplectic}.
These lose, however, the powerful machinery of extrapolation applied to standard NCSM calculations. 

Finally, rather than being `smarter' in our physics, one can ride a current trend and hand over insights to the computer, with
novel extrapolations  using machine learning \citep{negoita2018deep}.  The initial results are impressive, and it remains to see how 
widespread such techniques can be applied.




\Acknowledgements

This material is based upon work supported by the U.S. Department of Energy, Office of Science, Office of Nuclear Physics, 
under Award Number  DE-FG02-03ER41272.

\bibliographystyle{unsrtnat}
\bibliography{johnsonmaster}




\end{document}